\documentclass[%
 onecolumn,
 superscriptaddress,
 nofootinbib,
 amsmath,
 aps,
 prb,
 10pt
]{revtex4-2}

\usepackage[pdftex]{graphicx, color}
\usepackage[pdftex,
            colorlinks=true,
            citecolor=blue,
            linkcolor=blue]{hyperref}
\usepackage{newtxtext, newtxmath}
\usepackage{bm}
\begin{document}

\title{Topological gapless points in superconductors: From the viewpoint of symmetry}

\author{Shuntaro Sumita}
\email[]{shuntaro.sumita@riken.jp}
\affiliation{%
 Condensed Matter Theory Laboratory, RIKEN CPR, Wako, Saitama 351-0198, Japan
}%

\author{Youichi Yanase}
\affiliation{%
 Department of Physics, Kyoto University, Kyoto 606-8502, Japan
}%
\affiliation{%
 Institute for Molecular Science, 38 Nishigo-Naka, Myodaiji, Okazaki, Aichi 444-8585, Japan
}%

\date{\today}

\begin{abstract}%
 Searching for topological insulators/superconductors is a central subject in recent condensed matter physics.
 As a theoretical aspect, various classification methods of symmetry-protected topological phases have been developed, where the topology of a gapped Hamiltonian is investigated from the viewpoint of its onsite/crystal symmetry.
 On the other hand, topological physics also appears in semimetals, whose gapless points can be characterized by topological invariants.
 Stimulated by this background, we shed light on the topology of nodal superconductors.
 In this paper, we review our modern topological classification theory of superconducting gap nodes in terms of symmetry.
 The classification method elucidates nontrivial gap structures arising from nonsymmorphic symmetry or angular momentum, which cannot be predicted by a conventional theory.
\end{abstract}

\maketitle

\section{Introduction}
\label{sec:introduction}
In recent condensed matter physics, topological phases of matter have attracted much attention~\cite{Hasan2010_review, Qi2011_review, Chiu2016_review}.
So far many researchers have energetically proposed various topological materials, where the wavefunction has nontrivial topology characterized by a topological invariant.
A topological invariant is well-defined for a gapped Hamiltonian, i.e., an insulator or a fully gapped superconductor.
Over the past decade, theoretical studies have shown that topological numbers can be classified by the symmetry and dimensionality of the system.
In the early stages of the studies, classification theory based on \textit{onsite} symmetries, namely time-reversal symmetry (TRS), particle--hole symmetry (PHS), and chiral symmetry (CS), was developed~\cite{Schnyder2008, Kitaev2009, Ryu2010}.
The three symmetries categorize any system into ten types, which are known as Altland--Zirnbauer (AZ) classes~\cite{Altland1997, Zirnbauer1996}.
Based on the categorization, Refs.~\cite{Schnyder2008, Kitaev2009, Ryu2010} have suggested a \textit{topological periodic table}.
One can identify, by using the table, a topological number for a (gapped) system of interest, when the AZ class and dimension of the system are determined.
On the other hand, more recent studies have intensively investigated a classification problem under \textit{crystal} symmetry as well as onsite symmetry.
The first trigger for such studies was a suggestion of topological crystalline insulators by Fu~\cite{Fu2011}.%
\footnote{It has been revealed that Fu's model represents a kind of fragile topological insulator~\cite{Alexandradinata2020}.}
Since then, many researchers have challenged the exhaustive classification of topological invariants based on (magnetic) point/space group symmetry, by developing various methods: band structure combinatorics~\cite{Slager2012, Kruthoff2017, Bouhon2021}, topological quantum chemistry~\cite{Bradlyn2017, Bradlyn2018, Cano2018, Vergniory2019, Vergniory2021_arXiv, Cano2021}, symmetry-based indicators~\cite{Po2017, Watanabe2018, Khalaf2018, Ono2018, Ono2019, Ono2020, Ono2021}, Atiyah--Hirzebruch spectral sequence~\cite{Shiozaki2018_arXiv1, Shiozaki2018_arXiv2, Okuma2019}, etc.

Although the topology of a wavefunction can be defined only for a gapped Hamiltonian, topological invariants are useful even for \textit{gapless} systems.
Indeed, Weyl semimetals possess gapless points (Weyl points) characterized by a 2D invariant called a Chern number.
In other words, when the Chern number on a certain 2D slice in the 3D Brillouin zone (BZ) is different from that on another slice, there must be Weyl point(s) between the slices.
Furthermore, Refs.~\cite{Wang2012, BJYang2014} have suggested an additional topological number accompanied by the presence of rotation symmetry, which protects a Dirac point on the rotation axis in the BZ.
The Dirac point is stable as long as the relevant rotation symmetry is preserved.
These findings provide the following two insights.
First, almost all gapless points in semimetals are characterized by (weak) topological invariants.
Second, the presence or absence of the invariants determines the stability of the gapless points against fluctuations.

On the other hand, some superconductors also have gapless points, called superconducting nodes.
Since the momentum dependence of the superconducting gap is closely related to the symmetry of superconductivity and the pairing mechanism, it is important to investigate the structure and the stability of superconducting nodes.
Although previous theories have pointed out some examples of topologically protected superconducting nodes~\cite{Meng2012, Sau2012, SAYang2014, Volovik2017, Kozii2016, Venderbos2016, Goswami2015, Fischer2014, Daido2016, Yanase2017, Yanase2016}, comprehensive understanding of the relationships between topology and nodes has not been obtained.
Moreover, recently developed group-theoretical classification theories of superconducting gap structures have discovered unusual nodes due to nonsymmorphic crystal symmetry in multi-sublattice superconductors~\cite{Norman1995, Micklitz2009, Kobayashi2016, Yanase2016, Nomoto2016_PRL, Nomoto2017, Micklitz2017_PRB, Micklitz2017_PRL, Sumita2017, Sumita2018}, or higher-spin states in multi-orbital ones~\cite{Sumita2018, Nomoto2016_PRB, Wan2009, Brydon2016, Agterberg2017, Timm2017, Savary2017, Boettcher2018, Kim2018, Venderbos2018}.
The nontrivial nodal structures cannot be predicted by a conventional classification theory of order parameters based on point-group symmetry~\cite{Volovik1984, Volovik1985, Anderson1984, Blount1985, Ueda1985, Sigrist-Ueda}, for the following two reasons:
(1) the \textit{order-parameter} classification is sometimes incompatible with the superconducting \textit{gap} structure (i.e., an excitation energy in the Bogoliubov spectrum), and
(2) the theory does not take into account space group symmetry and higher-spin states characteristic of superconductors with multi-degrees of freedom.
Therefore, a new classification theory of superconducting nodes, relevant to topology and resolving the above two problems, was needed.

Because of the above background, we have actually constructed such a modern classification theory and discovered unconventional superconducting nodes protected by crystal symmetry and topology~\cite{Kobayashi2018, Sumita2019, Sumita2020_kotaibutsuri, Sumita2021_book}.
In this paper, we review the method (Sect.~\ref{sec:method}) and the result of the topological crystalline superconducting nodes (Sect.~\ref{sec:result}).
Finally, a brief summary and related topics are given in Sect.~\ref{sec:summary}.
The content in the paper overlaps with the Japanese review article~\cite{Sumita2020_kotaibutsuri} written by one of the authors and Shingo Kobayashi.

\section{Method}
\label{sec:method}
In this section, we explain the modern classification method of superconducting gap nodes, using group theory and topological arguments.
First, for the avoidance of confusion, in Sect.~\ref{sec:preliminary} we introduce many terminologies and notations, that are used throughout the paper.
Next, in Sect.~\ref{sec:gap_classification_topology}, we construct the topological classification theory of nodes on the high-symmetry points by using the Wigner criteria and the orthogonality test~\cite{Kobayashi2014, Kobayashi2016, Kobayashi2018, Sumita2019, Sumita2020_kotaibutsuri, Sumita2021_book}.

\subsection{Preliminary}
\label{sec:preliminary}
In this section, we define many terminologies and notations of finite-group representation theory, in preparation for the main topological classification (Sect.~\ref{sec:gap_classification_topology}).
In all the discussions below, we focus on \textit{centrosymmetric} superconductors that possess inversion symmetry (IS) $\mathcal{I} = \{I | \bm{0}\}$.\footnote{The notation $g = \{p_g | \bm{t}_g\}$ is a conventional Seitz space group symbol with a point-group operation $p_g$ and a translation $\bm{t}_g$.}
When a superconductor of interest is paramagnetic (PM) and antiferromagnetic (AFM), it has TRS $\mathcal{T} = \{T | \bm{t}_\mathcal{T}\}$, while the TRS is broken for a ferromagnetic (FM) superconductor.
Note that $\bm{t}_\mathcal{T}$ represents a nonprimitive translation for AFM superconductors,%
\footnote{Even when $\bm{t}_\mathcal{T}$ is nonzero, we treat the symmetry $\{T | \bm{t}_\mathcal{T}\}$ as a kind of TRS, although one often calls it a magnetic translation symmetry.}
while it is zero for PM superconductors.
Furthermore, we restrict our target to \textit{spin--orbit-coupled} superconductors, although the classification theory introduced below is straightforwardly applicable to spinless systems.

First, let $G$ and $M$ be unitary and magnetic space groups of the system of interest, respectively.
$M$ and $G$ include a translation group $\mathbb{T}$, which is determined by the Bravais lattice.
Due to the above assumption, the space group includes the IS $\mathcal{I}$.
When we consider PM and AFM superconductors, $M$ is equal to $G + \mathcal{T} G$, while $M = G$ for FM superconductors.
In order to construct the new classification theory resolving the problems listed in the introduction, we need to fix a specific $\bm{k}$ point in the BZ.
Since we are particularly interested in a classification problem of the superconducting gap structure on high-symmetry $\bm{k}$ points (mirror planes and rotation axes), we focus on the operations in $G$ ($M$) leaving the $\bm{k}$ points invariant modulo a reciprocal lattice vector, generating a subgroup of $G$ ($M$).
The subgroup $\mathcal{G}^{\bm{k}} < G$ ($\mathcal{M}^{\bm{k}} < M$) is called a (magnetic) little group.
The factor group of the (magnetic) little group by the translation group $\bar{\mathcal{G}}^{\bm{k}} = \mathcal{G}^{\bm{k}} / \mathbb{T}$ ($\bar{\mathcal{M}}^{\bm{k}} = \mathcal{M}^{\bm{k}} / \mathbb{T}$) is called a (magnetic) little cogroup.
The (magnetic) little cogroup is isomorphic to a certain (magnetic) point group.
Indeed, the little cogroup satisfies
\begin{align}
 \bar{\mathcal{G}}^{\bm{k}} &\cong C_s, \\
 \intertext{for mirror planes, and}
 \bar{\mathcal{G}}^{\bm{k}} &\cong C_n \ \text{or} \ C_{nv},
\end{align}
for $n$-fold rotation axes ($n = 2$, $3$, $4$, and $6$).
Symmetry operations in the point groups $C_s$ and $C_n$ are explicitly represented by
\begin{align}
 C_s &= \{E, \mathcal{M}_z\}, \\
 C_n &= \{E, (\mathcal{C}_n)^1, \dots, (\mathcal{C}_n)^{n-1}\},
\end{align}
where $E$, $\mathcal{M}_z$, and $\mathcal{C}_n$ are an identity operator, a mirror operator (perpendicular to the $z$ axis), and an $n$-fold rotation operator, respectively.
The point group $C_{nv}$ is generated by the rotation $\mathcal{C}_n$ and a vertical mirror operator, whose mirror plane includes the rotation axis.
In each case, the magnetic little cogroup is given by
\begin{align}
 \bar{\mathcal{M}}^{\bm{k}} &\cong
 \begin{cases}
  C_s, & \text{in FM superconductors}, \\
  C_s + \mathcal{T I} C_s, & \text{in PM or AFM superconductors},
 \end{cases} \\
 \intertext{for mirror planes, and}
 \bar{\mathcal{M}}^{\bm{k}} &\cong
 \begin{cases}
  C_{n(v)}, & \text{in FM superconductors}, \\
  C_{n(v)} + \mathcal{T I} C_{n(v)}, & \text{in PM or AFM superconductors},
 \end{cases}
 \label{eq:magnetic_little_cogroup_rotation}
\end{align}
for $n$-fold rotation axes ($n = 2$, $3$, $4$, and $6$).

Next, let $\lambda^{\bm{k}}_\alpha$ be a double-valued irreducible representation (IR) of the magnetic little group $\mathcal{M}^{\bm{k}}$, which represents a normal Bloch state with the crystal momentum $\bm{k}$:
\begin{equation}
 m c_{\alpha i}^\dagger(\bm{k}) m^{- 1} = \sum_{j} c_{\alpha j}^\dagger(\bm{k}) [\lambda^{\bm{k}}_\alpha(m)]_{j i}, \quad m \in \mathcal{M}^{\bm{k}}.
 \label{eq:Bloch_trsf}
\end{equation}
$c_{\alpha i}^\dagger(\bm{k})$ is a creation operator of the Bloch state with a crystal momentum $\bm{k}$.
$\alpha$ is a label of the (double-valued) IR, which corresponds to the total angular momentum of the Bloch state $j_z = \pm 1 / 2, \pm 3 / 2, \dotsc$ in spin--orbit coupled systems.%
\footnote{Strictly speaking, $\alpha$ is an IR of the \textit{finite} group $\bar{\mathcal{M}}^{\bm{k}} = \mathcal{M}^{\bm{k}} / \mathbb{T}$, while $j_z$ is a basis of the \textit{continuous} (rotation) group. Therefore there is no one-to-one correspondence between $\alpha$ and $j_z$. For example, an IR $\alpha = 1 / 2$ of a point group $C_{2v}$ includes all normal Bloch states with half-integer total angular momentum $j_z = \pm 1 / 2, \pm 3 / 2, \pm 5 / 2, \dotsc$.}
$\lambda^{\bm{k}}_\alpha$ is called a small corepresentation.
When we consider the momentum $\bm{k}$ on a twofold rotation axis in a PM superconductor, for instance, $\bar{\mathcal{M}}^{\bm{k}}$ is given by $\{E, \mathcal{C}_2, \mathcal{T I}, \mathcal{T M}_z\}$ [see Eq.~\eqref{eq:magnetic_little_cogroup_rotation}].
Equation~\eqref{eq:Bloch_trsf} for $m = \mathcal{C}_2$ is then represented by
\begin{equation}
 \mathcal{C}_2 c_{1/2, i}^\dagger(\bm{k}) (\mathcal{C}_2)^{- 1} = \sum_{j=1,2} c_{1/2, j}^\dagger(\bm{k}) (-i \sigma_z)_{j i},
\end{equation}
where $\sigma_z$ is a Pauli matrix.
$c_{1/2, i}^\dagger(\bm{k})$ for $i = 1, 2$ describes the pseudo-spin-up and -down Bloch states, respectively.

Since $\mathcal{M}^{\bm{k}}$ is a semidirect product between the magnetic little cogroup $\bar{\mathcal{M}}^{\bm{k}}$ and the translation group $\mathbb{T}$, the small corepresentation satisfies
\begin{equation}
 \lambda^{\bm{k}}_\alpha(T_{\bm{R}}) = e^{- i \bm{k} \cdot \bm{R}}, \quad T_{\bm{R}} = \{E | \bm{R}\} \in \mathbb{T},
\end{equation}
where $\bm{R}$ is a Bravais lattice vector.
Furthermore, the algebra of $\lambda^{\bm{k}}_\alpha$ obeys a factor system $\{\omega_{\text{in}}(l_1, l_2)\} \in H^2(M, \text{U(1)}_\phi)$,%
\footnote{$H^2(M, \text{U(1)}_\phi)$ stands for the second cohomology class of the group $M$.}
which arises from internal degrees of freedom (e.g., a half-integer spin of electrons):%
\footnote{In spin-$1 / 2$ systems, for instance, a phase factor $-1$ accompanies two-time operations of TRS ($\mathcal{T}$), which means a factor system $\omega_{\text{in}}(\mathcal{T}, \mathcal{T}) = -1$.
Supposing that $\bm{k}_0$ is a time-reversal invariant momentum, therefore, the representation matrix obeys the algebra $\lambda^{\bm{k}_0}(\mathcal{T}) \lambda^{\bm{k}_0}(\mathcal{T})^* = - \lambda^{\bm{k}_0}(E)$.}
\begin{equation}
 \omega_{\text{in}}(m_1, m_2) \lambda^{\bm{k}}_\alpha(m_1 m_2) =
 \begin{cases}
  \lambda^{\bm{k}}_\alpha(m_1) \lambda^{\bm{k}}_\alpha(m_2), & \phi(m_1) = 1, \\
  \lambda^{\bm{k}}_\alpha(m_1) \lambda^{\bm{k}}_\alpha(m_2)^*, & \phi(m_1) = - 1,
 \end{cases}
 \label{eq:small_corepresentation_product}
\end{equation}
where $\phi\colon \mathcal{M} \to \mathbb{Z}_2$ is defined by
\begin{equation}
 \phi(m) =
 \begin{cases}
  +1, & m: \text{unitary}, \\
  -1, & m: \text{antiunitary}.
 \end{cases}
\end{equation}
Since the projective representation $\lambda^{\bm{k}}_\alpha$ complies with the associativity law, the factor system satisfies the following 2-cocycle condition:
\begin{equation}
 \omega_{\text{in}}(m_1, m_2) \omega_{\text{in}}(m_1 m_2, m_3) = \omega_{\text{in}}(m_1, m_2 m_3) \omega_{\text{in}}(m_2, m_3)^{\phi(m_1)}.
\end{equation}

For easy treatment of the small corepresentation $\lambda^{\bm{k}}_\alpha$ of the magnetic little group $\mathcal{M}^{\bm{k}}$ with an \textit{infinite} number of elements, we instead consider the \textit{finite} group (magnetic little cogroup) $\bar{\mathcal{M}}^{\bm{k}}$.
A representation matrix $\bar{\lambda}^{\bm{k}}_\alpha$ on $\bar{\mathcal{M}}^{\bm{k}}$ corresponding to $\lambda^{\bm{k}}_\alpha$ of $\mathcal{M}^{\bm{k}}$ is introduced by
\begin{equation}
 \lambda^{\bm{k}}_\alpha(m) = e^{- i \bm{k} \cdot \bm{\tau}_m} \bar{\lambda}^{\bm{k}}_\alpha(\bar{m}),
 \label{eq:small_corepresentation_finite}
\end{equation}
where $\bar{m}$ is a representative in $\bar{\mathcal{M}}^{\bm{k}}$ for $m \in \mathcal{M}^{\bm{k}}$, and
\begin{equation}
 \bm{\tau}_m = \bm{t}_m - \bm{t}_{\bar{m}} \in \mathbb{T}
\end{equation}
is a Bravais lattice translation corresponding to $m$.
Substituting Eq.~\eqref{eq:small_corepresentation_finite} into Eq.~\eqref{eq:small_corepresentation_product}, we obtain
\begin{gather}
 \omega_{\text{in}}(m_1, m_2) e^{- i \bm{k} \cdot \bm{\nu}(m_1, m_2)} \bar{\lambda}^{\bm{k}}_\alpha(\overline{m_1 m_2}) =
 \begin{cases}
  \bar{\lambda}^{\bm{k}}_\alpha(\overline{m_1}) \bar{\lambda}^{\bm{k}}_\alpha(\overline{m_2}), & \phi(m_1) = 1, \\
  \bar{\lambda}^{\bm{k}}_\alpha(\overline{m_1}) \bar{\lambda}^{\bm{k}}_\alpha(\overline{m_2})^*, & \phi(m_1) = - 1,
 \end{cases}
 \label{eq:small_corepresentation_product_mpg} \\
 \bm{\nu}(m_1, m_2) = \bm{\tau}_{m_1 m_2} - \bm{\tau}_{m_1} - p_{m_1} \bm{\tau}_{m_2} = - [\bm{t}_{\overline{m_1 m_2}} - \bm{t}_{\overline{m_1}} - p_{m_1} \bm{t}_{\overline{m_2}}],
\end{gather}
where $p_{m_1}$ is the point-group part of the operator $m_1$.
In Eq.~\eqref{eq:small_corepresentation_product_mpg}, the nonsymmorphic part of the factor system%
\footnote{%
The factor system also satisfies the 2-cocycle condition:
\begin{equation*}
 \omega_{\text{ns}}^{\bm{k}}(m_1, m_2) \omega_{\text{ns}}^{\bm{k}}(m_1 m_2, m_3) = \omega_{\text{ns}}^{\bm{k}}(m_1, m_2 m_3) \omega_{\text{ns}}^{\bm{k}}(m_2, m_3)^{\phi(m_1)}.
\end{equation*}
}
is described by a Bravais lattice translation $\bm{\nu}(m_1, m_2)$, namely
\begin{equation}
 \omega_{\text{ns}}^{\bm{k}}(m_1, m_2) = \omega_{\text{ns}}^{\bm{k}}(\overline{m_1}, \overline{m_2}) = e^{- i \bm{k} \cdot \bm{\nu}(m_1, m_2)}.
 \label{eq:factor_system_ns}
\end{equation}
Thus, the small corepresentation $\lambda^{\bm{k}}_\alpha$ of the infinite group $\mathcal{M}^{\bm{k}}$ falls into the representation $\bar{\lambda}^{\bm{k}}_\alpha$ of the finite group $\bar{\mathcal{M}}^{\bm{k}}$, by introducing an appropriate factor system $\{\omega^{\bm{k}}(m_1, m_2) = \omega_{\text{in}}(m_1, m_2) \omega_{\text{ns}}^{\bm{k}}(m_1, m_2)\}$.

In a similar way, a projective representation $\bar{\gamma}^{\bm{k}}_\alpha$ of the unitary little cogroup $\bar{\mathcal{G}}^{\bm{k}}$, which corresponds to a small representation $\gamma^{\bm{k}}_\alpha$ of the little group $\mathcal{G}^{\bm{k}}$, is naturally introduced.
Supposing that the representation $\bar{\gamma}^{\bm{k}}_\alpha$ is concretely given, we can construct the corresponding corepresentation $\bar{\lambda}^{\bm{k}}_\alpha$.
For FM superconductors, $\bar{\lambda}^{\bm{k}}_\alpha$ is same as $\bar{\gamma}^{\bm{k}}_\alpha$ since $\bar{\mathcal{M}}^{\bm{k}} = \bar{\mathcal{G}}^{\bm{k}}$ due to TRS breaking.
When the superconductor is PM or AFM, on the other hand, we need to consider whether the degeneracy (dimension) of the representation increases for the nonunitary group $\bar{\mathcal{M}}^{\bm{k}} = \bar{\mathcal{G}}^{\bm{k}} + \mathfrak{T} \bar{\mathcal{G}}^{\bm{k}}$.
Here $\mathfrak{T} \equiv \mathcal{T I}$ is a pseudo-TRS operator preserved on any $\bm{k}$ point.
We can systematically solve the problem with the appropriate factor system~\cite{Bradley1968, Bradley-Cracknell}, by using a Wigner criterion (Herring test)~\cite{Wigner, Herring1937, Inui-Tanabe-Onodera, Bradley-Cracknell, Shiozaki2018_arXiv1}%
\footnote{The Wigner criterion for the \textit{pure} TRS indicates the presence or absence of a so-called Kramers degeneracy.}:
\begin{equation}
 W_\alpha^{\mathfrak{T}} \equiv \frac{1}{|\bar{\mathcal{G}}^{\bm{k}}|} \sum_{\bar{g} \in \bar{\mathcal{G}}^{\bm{k}}} \omega^{\bm{k}}(\mathfrak{T} \bar{g}, \mathfrak{T} \bar{g}) \chi[\bar{\gamma}^{\bm{k}}_\alpha(\overline{(\mathfrak{T} g)^2})] =
 \begin{cases}
  1, & \text{(a)}, \\
  -1, & \text{(b)}, \\
  0, & \text{(c)},
 \end{cases}
 \label{eq:Wigner_criterion_T}
\end{equation}
where $\chi$ is a character of the representation.
In the (b) and (c) cases, the degeneracy (dimension) of $\bar{\lambda}^{\bm{k}}_\alpha$ is twice that of $\bar{\gamma}^{\bm{k}}_\alpha$, while $\bar{\lambda}^{\bm{k}}_\alpha(\bar{g})$ gives the same representation as $\bar{\gamma}^{\bm{k}}_\alpha(\bar{g})$ for $\bar{g} \in \bar{\mathcal{G}}^{\bm{k}}$ in the (a) case.
For details, see Appendix~\ref{sec:Wigner_criterion}.

In the above discussion, we have introduced a lot of notations for groups and representations.
For the avoidance of confusion, we summarize the notations in Table~\ref{tab:notations}.

\begin{table}[tbp]
 \caption{Terminologies and notations with respect to group theory used in the paper. The first and second columns are associated with unitary groups, the third and fourth columns with nonunitary groups including antiunitary operators. In the table, we adopt the terminologies of Ref.~\cite{Bradley-Cracknell}. Adapted with permission from Ref.~\cite{Sumita2019}. Copyright \copyright{} 2019 by the American Physical Society.}
 \label{tab:notations}
 \begin{center}
  \small
  \begin{tabular}{lclcl} \hline\hline
   \multicolumn{1}{c}{Terminology} & Notation & \multicolumn{1}{c}{Terminology} & Notation & \multicolumn{1}{c}{Definition} \\ \hline
   Space group          & $G$                            & Magnetic space group    & $M$                             & Whole crystal symmetry of the system   \\
   Little group         & $\mathcal{G}^{\bm{k}}$         & Magnetic little group   & $\mathcal{M}^{\bm{k}}$          & Stabilizer of $\bm{k}$                 \\
   Small representation & $\gamma^{\bm{k}}_\alpha$       & Small corepresentation  & $\lambda^{\bm{k}}_\alpha$       & IR of (magnetic) little group          \\
   Little cogroup       & $\bar{\mathcal{G}}^{\bm{k}}$   & Magnetic little cogroup & $\bar{\mathcal{M}}^{\bm{k}}$    & Factor group of (magnetic) little group by $\mathbb{T}$ \\
   N/A                  & $\bar{\gamma}^{\bm{k}}_\alpha$ & N/A                     & $\bar{\lambda}^{\bm{k}}_\alpha$ & IR of (magnetic) little cogroup        \\ \hline\hline
  \end{tabular}
 \end{center}
\end{table}

\subsection{Topological classification of the superconducting gap}
\label{sec:gap_classification_topology}
In this section, we consider a topological classification method of the superconducting gap on high-symmetry points (mirror planes or rotation axes) in the BZ.
Before going into the details of the theory, let us give an overview of the general framework of the topological classification.
Our theory is based on the earlier classification method of topological insulators/superconductors~\cite{Schnyder2008, Kitaev2009, Ryu2010}.
Note that, in the usual sense, a topological insulator (superconductor) is an insulator (superconductor) where the wavefunctions of electrons (Bogoliubov quasiparticles) below the bandgap have nontrivial topology.
The topological invariant is defined for the (Bogoliubov--de Gennes, BdG) Hamiltonian on the BZ, which is represented by a $d$-dimensional torus.
As mentioned in the introduction, the topological periodic table (Table~\ref{tab:AZ_class}), which shows classification of the invariant based on the onsite symmetries, namely TRS ($\mathcal{T}$), PHS ($\mathcal{C}$), and CS ($\Gamma$), is widespread in current condensed matter physics~\cite{Schnyder2008, Kitaev2009, Ryu2010}.
Table~\ref{tab:AZ_class} includes the tenfold AZ classes~\cite{Zirnbauer1996, Altland1997}, which are classified by four types of algebraic structure: $0$, $\mathbb{Z}$, $2\mathbb{Z}$, and $\mathbb{Z}_2$.

\begin{table}[tbp]
 \caption{Correspondence between the AZ classes and the onsite symmetries: TRS ($\mathcal{T}$), PHS ($\mathcal{C}$) and CS ($\Gamma$)~\cite{Schnyder2008, Kitaev2009, Ryu2010}. 
 The first--third columns represent the presence or absence of the three symmetries. The absence of symmetries is denoted by ``$0$'', while the presence is denoted by either ``$+1$'' or ``$-1$'', depending on whether the antiunitary operator squares to $+1$ or $-1$.
 The fifth--eighth columns show topological classification results for a Hamiltonian defined on a $d$-dimensional torus for each AZ class. Classifications in the complex classes (A, AIII) and those in the real classes (AI, $\dotsc$, CI) are respectively periodic (Bott periodicity).}
 \label{tab:AZ_class}
 \begin{center}
  \begin{tabular}{cccccccc} \hline\hline
   $\mathcal{T}$ & $\mathcal{C}$ & $\Gamma$ & AZ class & $d = 0$ & $d = 1$ & $d = 2$ & $d = 3$ \\ \hline
   $0$  & $0$  & $0$ & A    & $\mathbb{Z}$   & $0$            & $\mathbb{Z}$   & $0$            \\
   $0$  & $0$  & $1$ & AIII & $0$            & $\mathbb{Z}$   & $0$            & $\mathbb{Z}$   \\[2mm]
   $+1$ & $0$  & $0$ & AI   & $\mathbb{Z}$   & $0$            & $0$            & $0$            \\
   $+1$ & $+1$ & $1$ & BDI  & $\mathbb{Z}_2$ & $\mathbb{Z}$   & $0$            & $0$            \\
   $0$  & $+1$ & $0$ & D    & $\mathbb{Z}_2$ & $\mathbb{Z}_2$ & $\mathbb{Z}$   & $0$            \\
   $-1$ & $+1$ & $1$ & DIII & $0$            & $\mathbb{Z}_2$ & $\mathbb{Z}_2$ & $\mathbb{Z}$   \\
   $-1$ & $0$  & $0$ & AII  & $2\mathbb{Z}$  & $0$            & $\mathbb{Z}_2$ & $\mathbb{Z}_2$ \\
   $-1$ & $-1$ & $1$ & CII  & $0$            & $2\mathbb{Z}$  & $0$            & $\mathbb{Z}_2$ \\
   $0$  & $-1$ & $0$ & C    & $0$            & $0$            & $2\mathbb{Z}$  & $0$            \\
   $+1$ & $-1$ & $1$ & CI   & $0$            & $0$            & $0$            & $2\mathbb{Z}$  \\ \hline\hline
  \end{tabular}
 \end{center}
\end{table}

Now we apply the above classification theory of topological insulators/superconductors~\cite{Schnyder2008, Kitaev2009, Ryu2010} to the superconducting gap classification~\cite{Kobayashi2014, Kobayashi2018}.
However, a bandgap cannot exist in the whole $d$-dimensional BZ in nodal superconductors.
Therefore, we instead consider a $p$-dimensional sphere ($p < d$) surrounding the node.
$p + 1$ is called a codimension, which shows a difference between the dimension of the BZ and that of the node.
For example, when a (1D) line node exists in a 3D BZ, the codimension is $3 - 1 = 2$, which results in $p = 1$.

It is necessary to reconsider symmetry since the domain of a topological number is changed from the $d$-dimensional BZ to the $p$-dimensional sphere.
On the sphere, only symmetries preserving the position of the node are allowed.
In particular, the node should be fixed to a high-symmetry $\bm{k}$ point (mirror plane or rotational axis) in crystalline superconductors.
As mentioned in Sect.~\ref{sec:preliminary}, a unitary little cogroup on the high-symmetry $\bm{k}$ point is denoted by $\bar{\mathcal{G}}^{\bm{k}}$, and an IR of $\bar{\mathcal{G}}^{\bm{k}}$ by $\bar{\gamma}^{\bm{k}}_\alpha$, which corresponds to the normal Bloch state with the momentum $\bm{k}$.
Furthermore, although the TRS $\mathcal{T}$, the PHS $\mathcal{C}$, and the IS $\mathcal{I}$ themselves do not preserve the position of the node, the combined symmetries of them are allowed on any $\bm{k}$ point.
Therefore, important symmetries for our classification theory are pseudo-TRS $\mathfrak{T} \equiv \mathcal{T I}$, pseudo-PHS $\mathfrak{C} \equiv \mathcal{C I}$, and CS $\Gamma \equiv \mathcal{T C}$, all of which are preserved on any $\bm{k}$ point.
The stabilizer of the high-symmetry $\bm{k}$ point is thus represented by the following group:
\begin{equation}
 \bar{\mathfrak{G}}^{\bm{k}} = \bar{\mathcal{G}}^{\bm{k}} + \mathfrak{T} \bar{\mathcal{G}}^{\bm{k}} + \mathfrak{C} \bar{\mathcal{G}}^{\bm{k}} + \Gamma \bar{\mathcal{G}}^{\bm{k}}.
\end{equation}
A topological number on the high-symmetry point is defined by using the BdG Hamiltonian with the symmetry $\bar{\mathfrak{G}}^{\bm{k}}$.
Although the classification theory of topological insulators/superconductors~\cite{Schnyder2008, Kitaev2009, Ryu2010} is not directly applicable to the gap classification under the symmetry $\bar{\mathfrak{G}}^{\bm{k}}$, an \textit{effective} AZ (EAZ) class is instead defined for the BdG Hamiltonian on the $p$-dimensional sphere.
For this purpose, we execute the \textit{Wigner criteria}~\cite{Wigner, Herring1937, Inui-Tanabe-Onodera, Bradley-Cracknell, Shiozaki2018_arXiv1} for $\mathfrak{T}$ and $\mathfrak{C}$,
\begin{align}
 W_\alpha^{\mathfrak{T}} &\equiv \frac{1}{|\bar{\mathcal{G}}^{\bm{k}}|} \sum_{\bar{g} \in \bar{\mathcal{G}}^{\bm{k}}} \omega^{\bm{k}}(\mathfrak{T} \bar{g}, \mathfrak{T} \bar{g}) \chi[\bar{\gamma}^{\bm{k}}_\alpha(\overline{(\mathfrak{T} \bar{g})^2})] =
 \begin{cases}
  1, \\
  -1, \\
  0,
 \end{cases}
 \tag{\ref{eq:Wigner_criterion_T} revisited} \displaybreak[2] \\
 W_\alpha^{\mathfrak{C}} &\equiv \frac{1}{|\bar{\mathcal{G}}^{\bm{k}}|} \sum_{\bar{g} \in \bar{\mathcal{G}}^{\bm{k}}} \omega^{\bm{k}}(\mathfrak{C} \bar{g}, \mathfrak{C} \bar{g}) \chi[\bar{\gamma}^{\bm{k}}_\alpha(\overline{(\mathfrak{C} \bar{g})^2})] =
 \begin{cases}
  1, \\
  -1, \\
  0,
 \end{cases}
 \label{eq:Wigner_criterion_C}
 \intertext{and the \textit{orthogonality test}~\cite{Inui-Tanabe-Onodera, Shiozaki2018_arXiv1} for $\Gamma$:}
 W_\alpha^\Gamma &\equiv \frac{1}{|\bar{\mathcal{G}}^{\bm{k}}|} \sum_{\bar{g} \in \bar{\mathcal{G}}^{\bm{k}}} \frac{\omega^{\bm{k}}(\bar{g}, \Gamma)^*}{\omega^{\bm{k}}(\Gamma, \Gamma^{-1} \bar{g} \Gamma)^*} \chi[\bar{\gamma}^{\bm{k}}_\alpha(\overline{\Gamma^{-1} g \Gamma})^*] \chi[\bar{\gamma}^{\bm{k}}_\alpha(\bar{g})] =
 \begin{cases}
  1, \\
  0.
 \end{cases}
 \label{eq:orthogonality_test_G}
\end{align}
In the above tests, we investigate orthogonality between a basis set $\{c_{\alpha i}^\dagger(\bm{k})\}$ of the normal Bloch states and another one $\{a c_{\alpha i}^\dagger(\bm{k}) a^{-1}\}$ transformed by the symmetry $a$ ($= \mathfrak{T}$, $\mathfrak{C}$, or $\Gamma$).
From Eqs.~\eqref{eq:Wigner_criterion_T}, \eqref{eq:Wigner_criterion_C}, and \eqref{eq:orthogonality_test_G}, we obtain a set of three numbers $(W_\alpha^{\mathfrak{T}}, W_\alpha^{\mathfrak{C}}, W_\alpha^\Gamma)$, which identifies the EAZ symmetry class of the BdG Hamiltonian on the sphere by using the knowledge of $K$ theory~\cite{Zirnbauer1996, Altland1997, Shiozaki2018_arXiv1}.
Table~\ref{tab:effective_AZ_class} shows the correspondence between the set of $(W_\alpha^{\mathfrak{T}}, W_\alpha^{\mathfrak{C}}, W_\alpha^\Gamma)$ and the EAZ symmetry class.%
\footnote{$W_\alpha^{\mathfrak{T}} = W_\alpha^\Gamma = 0$ in FM superconductors, since both the pseudo-TRS ($\mathfrak{T}$) and the CS ($\Gamma$) are broken.}

\begin{table}[tbp]
 \caption{Correspondence table between the set of $(W_\alpha^{\mathfrak{T}}, W_\alpha^{\mathfrak{C}}, W_\alpha^\Gamma)$ and EAZ symmetry classes in centrosymmetric systems~\cite{Kobayashi2014, Zhao2016, Bzdusek2017, Sumita2019}. The fifth--seventh columns show topological classification results for the IR at the $\bm{k}$ point with the codimension $p + 1$. Note that the Bott periodicity of the real classes is reversed from that in Table~\ref{tab:AZ_class}, since $\mathfrak{T}$ and $\mathfrak{C}$ do not flip the momentum, unlike $\mathcal{T}$ and $\mathcal{C}$. Adapted with permission from Ref.~\cite{Sumita2019}. Copyright \copyright{} 2019 by the American Physical Society.}
 \label{tab:effective_AZ_class}
 \begin{center}
  \begin{tabular}{ccccccc} \hline\hline
   $\mathfrak{T}$ ($W_\alpha^{\mathfrak{T}}$) & $\mathfrak{C}$ ($W_\alpha^{\mathfrak{C}}$) & $\Gamma$ ($W_\alpha^\Gamma$) & EAZ class & $p = 0$ & $p = 1$ & $p = 2$ \\ \hline
   $0$  & $0$  & $0$ & A    & $\mathbb{Z}$   & $0$            & $\mathbb{Z}$   \\
   $0$  & $0$  & $1$ & AIII & $0$            & $\mathbb{Z}$   & $0$            \\[2mm]
   $+1$ & $0$  & $0$ & AI   & $\mathbb{Z}$   & $\mathbb{Z}_2$ & $\mathbb{Z}_2$ \\
   $+1$ & $+1$ & $1$ & BDI  & $\mathbb{Z}_2$ & $\mathbb{Z}_2$ & $0$            \\
   $0$  & $+1$ & $0$ & D    & $\mathbb{Z}_2$ & $0$            & $2\mathbb{Z}$  \\
   $-1$ & $+1$ & $1$ & DIII & $0$            & $2\mathbb{Z}$  & $0$            \\
   $-1$ & $0$  & $0$ & AII  & $2\mathbb{Z}$  & $0$            & $0$            \\
   $-1$ & $-1$ & $1$ & CII  & $0$            & $0$            & $0$            \\
   $0$  & $-1$ & $0$ & C    & $0$            & $0$            & $\mathbb{Z}$   \\
   $+1$ & $-1$ & $1$ & CI   & $0$            & $\mathbb{Z}$   & $\mathbb{Z}_2$ \\ \hline\hline
  \end{tabular}
 \end{center}
\end{table}

Furthermore, from the EAZ symmetry class, we can classify the IR at the (high-symmetry) $\bm{k}$ point into an algebra $0$, $\mathbb{Z}$, $2\mathbb{Z}$, or $\mathbb{Z}_2$ (Table~\ref{tab:effective_AZ_class}).
In this context, $(W_\alpha^{\mathfrak{T}}, W_\alpha^{\mathfrak{C}}, W_\alpha^\Gamma)$ gives a symmetry-based topological classification of the Hamiltonian at each $\bm{k}$ point on the plane (line).
When the plane (line) intersects a normal-state Fermi surface, a node composed of the intersection line (point) on the plane (line) hosts a codimension $1$, and therefore it is surrounded by a 0D sphere.
In other words, superconducting gap nodes on the plane (line) are classified by a 0D topological number (see $p = 0$ in Table~\ref{tab:effective_AZ_class}).
When the classification is nontrivial ($\mathbb{Z}$, $2\mathbb{Z}$, or $\mathbb{Z}_2$), the intersection leads to a node characterized by the topological invariant.
Otherwise, a gap opens at the intersection line (point).

\section{Results}
\label{sec:result}
In Sect.~\ref{sec:method}, the topological classification theory of superconducting gap nodes was introduced.
Although the conventional classification of order parameters can speculate on the presence or absence of nodes from the momentum dependence of basis functions~\cite{Volovik1984, Volovik1985, Anderson1984, Blount1985, Ueda1985, Sigrist-Ueda}, it sometimes fails to describe the correct gap structures~\cite{Norman1995, Micklitz2009, Kobayashi2016, Yanase2016, Nomoto2016_PRL, Nomoto2017, Micklitz2017_PRB, Micklitz2017_PRL, Sumita2017, Sumita2018, Nomoto2016_PRB, Wan2009, Brydon2016, Agterberg2017, Timm2017, Savary2017, Boettcher2018, Kim2018, Venderbos2018}.
On the other hand, our modern theory can exactly classify the gap structures by taking into account nonsymmorphic symmetry and higher-spin states.
Indeed, we have performed comprehensive classification of gap structures on high-symmetry planes~\cite{Kobayashi2016, Sumita2018, Kobayashi2018, Sumita2020_kotaibutsuri, Sumita2021_book} and lines~\cite{Sumita2018, Sumita2019, Sumita2020_kotaibutsuri, Sumita2021_book} in the BZ.
In this section, we explain various nontrivial results of topological crystalline nodes, whose topological protection is characteristic of crystalline systems, as elucidated by the studies.

\subsection{Classification on high-symmetry planes: nontrivial gap structures due to nonsymmorphic symmetry}
\label{sec:gap_classification_planes}
First, we introduce classification results on high-symmetry planes, where the factor system $\omega_{\text{ns}}^{\bm{k}}$ accompanied by \textit{nonsymmorphic symmetry} induces nontrivial gap structures.
In the following, for instance, let us consider a magnetic space group $M = P6_3/mmc1'$, which is one of the hexagonal and PM space groups including nonsymmorphic screw symmetry.%
\footnote{The space group is represented by Hermann--Mauguin notation. For details, see Refs.~\cite{InternationalTables, Inui-Tanabe-Onodera}.}
The PM space group represents the symmetry of the odd-parity superconductor UPt$_3$, which is known to possess unconventional line nodes protected by nonsymmorphic symmetry~\cite{Norman1995, Micklitz2009, Kobayashi2016, Yanase2016, Nomoto2016_PRL, Micklitz2017_PRB, Sumita2018}.

\subsubsection{Preliminary}
Since the magnetic space group $M$ includes a mirror operator $\mathcal{M}_z = \{M_z | \frac{\hat{z}}{2}\}$, we focus on mirror-invariant planes $k_z = 0$ and $k_z = \pi$ in the BZ.
The (magnetic) little groups on the planes are given by
\begin{align}
 \mathcal{G}^{\bm{k}} &= \mathbb{T} + \mathcal{M}_z \mathbb{T}, \\
 \mathcal{M}^{\bm{k}} &= \mathcal{G}^{\bm{k}} + \mathfrak{T} \mathcal{G}^{\bm{k}} = \mathbb{T} + \mathcal{M}_z \mathbb{T} + \mathfrak{T} \mathbb{T} + \mathcal{M}_z \mathfrak{T} \mathbb{T},
\end{align}
from which the (magnetic) little cogroups are obtained:
\begin{align}
 \bar{\mathcal{G}}^{\bm{k}} &= \mathcal{G}^{\bm{k}} / \mathbb{T} = \{E, \mathcal{M}_z\}, \label{eq:194_little_cogroup_Mz} \\
 \bar{\mathcal{M}}^{\bm{k}} &= \mathcal{M}^{\bm{k}} / \mathbb{T} = \{E, \mathcal{M}_z, \mathfrak{T}, \mathcal{M}_z \mathfrak{T}\}.
\end{align}
The factor system taking into account spin--orbit coupling and nonsymmorphic symmetry is as follows:
\begin{subequations}
 \label{eq:194_factor_system_Mz_T}
 \begin{gather}
  \omega_{\text{in}}(\mathcal{M}_z, \mathcal{M}_z) = - 1, \, \omega_{\text{in}}(\mathfrak{T}, \mathfrak{T}) = - 1, \, \omega_{\text{in}}(\mathcal{M}_z, \mathfrak{T}) = \omega_{\text{in}}(\mathfrak{T}, \mathcal{M}_z) = 1, \\
  \omega_{\text{ns}}^{\bm{k}}(\mathcal{M}_z, \mathcal{M}_z) = 1, \, \omega_{\text{ns}}^{\bm{k}}(\mathfrak{T} \mathcal{M}_z, \mathfrak{T} \mathcal{M}_z) = e^{i k_z},
 \end{gather}
\end{subequations}
where we note that $\omega_{\text{ns}}^{\bm{k}}$ depends on $k_z$ due to the existence of nonsymmorphic symmetry.

Considering $\omega^{\bm{k}}(\mathcal{M}_z, \mathcal{M}_z) = - 1$, we obtain projective IRs $\bar{\gamma}^{\bm{k}}_{\pm 1 / 2}$ of the little cogroup $\bar{\mathcal{G}}^{\bm{k}}$,
\begin{equation}
 \bar{\gamma}^{\bm{k}}_{\pm 1 / 2}(E) = 1, \, \bar{\gamma}^{\bm{k}}_{\pm 1 / 2}(\mathcal{M}_z) = \mp i, \label{eq:194_Bloch_rep_Mz}
\end{equation}
which correspond to (half-integer) spin-up and spin-down states, respectively.
For the spin-up one, a Wigner criterion~\cite{Wigner, Herring1937, Inui-Tanabe-Onodera, Bradley-Cracknell, Shiozaki2018_arXiv1} for the pseudo-TRS \eqref{eq:Wigner_criterion_T} is calculated as
\begin{align}
 W_{+ 1 / 2}^{\mathfrak{T}} &= \frac{1}{2} \sum_{\bar{g} \in \{E, \mathcal{M}_z\}} \omega^{\bm{k}}(\mathfrak{T} \bar{g}, \mathfrak{T} \bar{g}) \chi[\bar{\gamma}^{\bm{k}}_{+ 1 / 2}((\mathfrak{T} \bar{g})^2)] \notag \\
 &= \frac{1}{2} ( - 1 + e^{i k_z} ) =
 \begin{cases}
  0, & k_z = 0, \\
  -1, & k_z = \pi.
 \end{cases}
 \label{eq:194_Wigner_criterion_Mz_Au_T}
\end{align}
Therefore, a representation $\bar{\lambda}^{\bm{k}}_{+ 1 / 2}$ of the magnetic little cogroup $\bar{\mathcal{M}}^{\bm{k}}$ can be constructed as follows (see also Appendix~\ref{sec:Wigner_criterion}):
\begin{equation}
 \label{eq:194_magnetic_Bloch_rep_Mz}
 \begin{array}{ccccc} \hline\hline
  \bar{m} & E & \mathcal{M}_z & \mathfrak{T} & \mathcal{M}_z \mathfrak{T} \\ \hline
  \bar{\lambda}^{\bm{k}}_{+ 1 / 2}(\bar{m}) & \begin{pmatrix} 1 & 0 \\ 0 & 1 \end{pmatrix} & \begin{pmatrix} - i & 0 \\ 0 & + i e^{i k_z} \end{pmatrix} & \begin{pmatrix} 0 & - 1 \\ 1 & 0 \end{pmatrix} & \begin{pmatrix} 0 & i \\ i e^{i k_z} & 0 \end{pmatrix} \\ \hline\hline
 \end{array}
\end{equation}
The dimension two of $\bar{\lambda}^{\bm{k}}_{+ 1 / 2}$ physically indicates degeneracy between the spin-up and spin-down states because of the pseudo-TRS.
Even when we start from the spin-down state, therefore, a projective representation equivalent to Eq.~\eqref{eq:194_magnetic_Bloch_rep_Mz} is obtained.
From the above result, a small corepresentation for an element $m \in \mathcal{M}^{\bm{k}}$ is given by
\begin{equation}
 \lambda^{\bm{k}}_{1 / 2}(m) = e^{- i \bm{k} \cdot \bm{R}} \bar{\lambda}^{\bm{k}}_{1 / 2}(\bar{m}),
\end{equation}
where $m = \bar{m} T_{\bm{R}}$, $\bar{m} \in \bar{\mathcal{M}}^{\bm{k}}$, and $T_{\bm{R}} = \{E | \bm{R}\} \in \mathbb{T}$.

\subsubsection{Topological gap classification}
Now let us consider the topological classification of the superconducting gap structures (Sect.~\ref{sec:gap_classification_topology}) in this system.
Before going to the main issue, we revisit a general result obtained by classification under IS; we temporarily forget the presence of the mirror symmetry in the paragraph.
A set of symmetries protecting a superconducting node, namely the pseudo-TRS $\mathfrak{T}$, the pseudo-PHS $\mathfrak{C}$, and the CS $\Gamma$, is classified by the ten EAZ classes in Table~\ref{tab:effective_AZ_class}.
The class on a general $\bm{k}$ point is D or DIII for even-parity superconductivity, and C or CII for odd-parity.
In particular, let us focus on the classification of line nodes ($p = 1$) in 3D odd-parity superconductors, which shows no topological protection of the nodes for both classes C and CII.
This fact is known as Blount's theorem, claiming the absence of line nodes in (spin--orbit-coupled) odd-parity superconductors~\cite{Blount1985, Kobayashi2014}.%
\footnote{As shown in the later discussions, Blount's theorem breaks down in some nonsymmorphic superconductors, since only point-group symmetry is considered in his theory.}
According to $p = 0$ for the EAZ class D in Table~\ref{tab:effective_AZ_class}, furthermore, we can predict the existence of a surface node characterized by a $
\mathbb{Z}_2$ invariant in 3D even-parity chiral superconductors.
The node is nothing but a \textit{Bogoliubov Fermi surface} suggested by recent theoretical studies~\cite{Agterberg2017, Timm2017, Brydon2018}.

From the above general results, an additional symmetry is essential for a stable line node in odd-parity superconductivity.
As mentioned above, indeed, Refs.~\cite{Norman1995, Micklitz2009, Kobayashi2016, Yanase2016, Nomoto2016_PRL, Micklitz2017_PRB, Sumita2018} have proposed a counter-example of Blount's theorem, i.e. the existence of line nodes in the odd-parity superconductor UPt$_3$, by taking into account the effect of nonsymmorphic symmetry.
In the following, we reproduce the counter-example from the viewpoint of the topological classification theory~\cite{Kobayashi2016, Kobayashi2018}.
For this purpose, let us identify the EAZ class on the mirror-invariant planes $k_z = 0$ and $k_z = \pi$, supposing a Cooper pair wavefunction belonging to the $A_u$ IR of the point group $C_{2h}$.%
\footnote{A candidate symmetry of a superconducting order parameter in UPt$_3$ is the IR $E_{2u}$ of the point group $D_{6h}$, which corresponds to the $A_u$ IR of the subgroup $C_{2h}$.}
The $A_u$ pairing symmetry results in the following factor systems:
\begin{subequations}
 \label{eq:194_factor_system_Mz_Au_C}
 \begin{gather}
  \omega_{\text{in}}(\mathfrak{C}, \mathfrak{C}) = - 1, \, \omega_{\text{in}}(\mathfrak{C}, \mathcal{M}_z) = - \omega_{\text{in}}(\mathcal{M}_z, \mathfrak{C}) = 1, \\
  \omega_{\text{ns}}^{\bm{k}}(\mathfrak{C} \mathcal{M}_z, \mathfrak{C} \mathcal{M}_z) = e^{i k_z},
 \end{gather}
\end{subequations}
and therefore
\begin{subequations}
 \label{eq:194_factor_system_Mz_Au_G}
 \begin{gather}
  \omega_{\text{in}}(\Gamma, \mathcal{M}_z) = - \omega_{\text{in}}(\mathcal{M}_z, \Gamma), \\
  \omega_{\text{ns}}^{\bm{k}}(\Gamma, \mathcal{M}_z) = \omega_{\text{ns}}^{\bm{k}}(\mathcal{M}_z, \Gamma) = 1.
 \end{gather}
\end{subequations}
Using Eqs.~\eqref{eq:194_factor_system_Mz_Au_C} and \eqref{eq:194_factor_system_Mz_Au_G}, we apply the topological classification to the representation $\bar{\gamma}^{\bm{k}}_{+ 1 / 2}$ of the normal Bloch state.
The Wigner criteria for $\mathfrak{T}$ and $\mathfrak{C}$ \eqref{eq:Wigner_criterion_T}, \eqref{eq:Wigner_criterion_C}, and an orthogonality test for $\Gamma$ \eqref{eq:orthogonality_test_G} are calculated as
\begin{align}
 W_{+ 1 / 2}^{\mathfrak{T}} &=
 \begin{cases}
  0, & k_z = 0, \\
  -1, & k_z = \pi,
 \end{cases}
 \tag{\ref{eq:194_Wigner_criterion_Mz_Au_T} revisited} \displaybreak[2] \\
 W_{+ 1 / 2}^{\mathfrak{C}} &= \frac{1}{2} \left\{ \omega^{\bm{k}}(\mathfrak{C}, \mathfrak{C}) \chi[\bar{\gamma}^{\bm{k}}_{+ 1 / 2}(\overline{\mathfrak{C}^2})] + \omega^{\bm{k}}(\mathfrak{C} \mathcal{M}_z, \mathfrak{C} \mathcal{M}_z) \chi[\bar{\gamma}^{\bm{k}}_{+ 1 / 2}(\overline{(\mathfrak{C} \mathcal{M}_z)^2})] \right\} \notag \\
 &= \frac{1}{2} ( - 1 - e^{i k_z} ) =
 \begin{cases}
  -1, & k_z = 0, \\
  0, & k_z = \pi,
 \end{cases} \label{eq:194_Wigner_criterion_Mz_Au_C} \displaybreak[2] \\
 W_{+ 1 / 2}^\Gamma &= \frac{1}{2} \left\{ \frac{\omega^{\bm{k}}(E, \Gamma)}{\omega(\Gamma, \Gamma^{-1} E \Gamma)} \chi[\bar{\gamma}^{\bm{k}}_{+ 1 / 2}(E)^*] \chi[\bar{\gamma}^{\bm{k}}_{+ 1 / 2}(E)] \right. \notag \\
 &\quad \left. + \frac{\omega^{\bm{k}}(\mathcal{M}_z, \Gamma)}{\omega(\Gamma, \Gamma^{-1} \mathcal{M}_z \Gamma)} \chi[\bar{\gamma}^{\bm{k}}_{+ 1 / 2}(\overline{\Gamma^{-1} \mathcal{M}_z \Gamma})^*] \chi[\bar{\gamma}^{\bm{k}}_{+ 1 / 2}(\mathcal{M}_z)] \right\} \notag \\
 &= \frac{1}{2} ( 1 - 1 ) = 0, \quad (k_z = 0, \pi). \label{eq:194_orthogonality_test_Mz_Au_G}
\end{align}
According to Table~\ref{tab:AZ_class}, therefore, a classifying space of the BdG Hamiltonian on the $k_z = 0$ ($k_z = \pi$) plane is the EAZ class C (AII).
Topological classification of a line node on the 2D plane ($p = 0$) is $0$ for the class C, and $2\mathbb{Z}$ for the class AII.
This indicates that superconducting gap structures are different between the basal plane $k_z = 0$ and the zone face $k_z = \pi$; the $A_u$ pair wavefunction opens its gap on $k_z = 0$, while a line node characterized by a $2\mathbb{Z}$ invariant emerges on $k_z = \pi$.
These results are entirely consistent with those of the group-theoretical arguments~\cite{Norman1995, Micklitz2009, Kobayashi2016, Yanase2016, Nomoto2016_PRL, Micklitz2017_PRB, Sumita2018}.

\begin{figure}[tbp]
 \centering
 \includegraphics[width=14cm, clip]{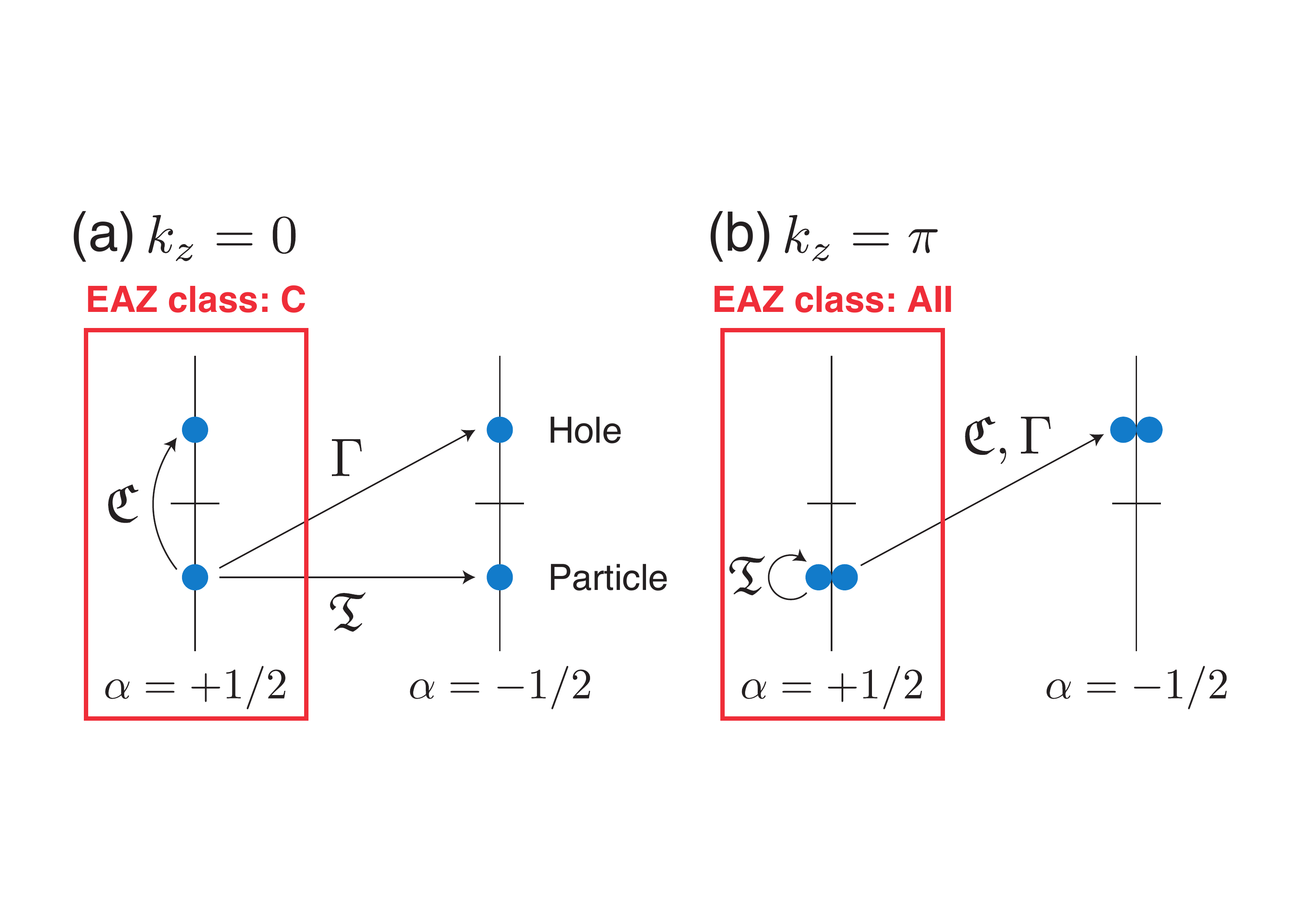}
 \caption{Schematic pictures of the BdG Hamiltonian with an $A_u$ pair wavefunction on (a) $k_z = 0$ and (b) $k_z = \pi$. The red frames in (a) and (b) belong to EAZ classes C and AII, respectively.}
 \label{fig:effective_AZ_class_Mz_Au}
\end{figure}

Now we consider the meaning of the classification results by the EAZ classes.
Figure~\ref{fig:effective_AZ_class_Mz_Au} schematically shows the structure of the BdG Hamiltonian with the $A_u$ Cooper pair wavefunction.
On the mirror-invariant planes ($k_z = 0, \pi$), the Hamiltonian can be block-diagonalized into the two IRs $\alpha = \pm 1 / 2$, due to the presence of mirror symmetry.
Let us first see the classification on the basal plane $k_z = 0$ [Fig.~\ref{fig:effective_AZ_class_Mz_Au}(a)].
In the above discussion, we started from the representation matrix $\bar{\gamma}^{\bm{k}}_{+ 1 / 2}$ of the IR $\alpha = + 1 / 2$, which corresponds to the spin-up normal Bloch state [the lower left particle in Fig.~\ref{fig:effective_AZ_class_Mz_Au}(a)].
The Wigner criterion for the pseudo-TRS operator, namely $W_{+ 1 / 2}^{\mathfrak{T}} = 0$ \eqref{eq:194_Wigner_criterion_Mz_Au_T}, indicates 
that a basis of an IR \textit{nonequivalent} to $\alpha = + 1 / 2$ is generated by $\mathfrak{T}$.
Therefore the lower left particle in Fig.~\ref{fig:effective_AZ_class_Mz_Au}(a) is mapped by $\mathfrak{T}$ to the lower right particle, which belongs to the other IR $\alpha = - 1 / 2$.
Similarly, the lower left particle is mapped by $\Gamma$ to the upper right hole, because of the orthogonality test $W_{+ 1 / 2}^{\Gamma} = 0$ \eqref{eq:194_orthogonality_test_Mz_Au_G}.
On the other hand, since the Wigner criterion for the pseudo-PHS operator leads to $W_{+ 1 / 2}^{\mathfrak{C}} = -1$ \eqref{eq:194_Wigner_criterion_Mz_Au_C}, $\mathfrak{C}$ gives a basis belonging to the \textit{equivalent} IR, which is linearly independent of the original basis.
The lower left particle in Fig.~\ref{fig:effective_AZ_class_Mz_Au}(a) is thus mapped by $\mathfrak{C}$ to the upper left hole in the same IR $\alpha = + 1 / 2$.
For the above reason, the Hamiltonian in the $\alpha = + 1 / 2$ space [the red frame in Fig.~\ref{fig:effective_AZ_class_Mz_Au}(a)] has only the pseudo-PHS $\mathfrak{C}$ with $\mathfrak{C}^2 = - E$, which indicates that the Hamiltonian block is classified into the AZ class C~\cite{Zirnbauer1996, Altland1997}.

However, the situation is different on the zone face $k_z = \pi$ [Fig.~\ref{fig:effective_AZ_class_Mz_Au}(b)].
In this case, the lower left particle in Fig.~\ref{fig:effective_AZ_class_Mz_Au}(b) is degenerated by $\mathfrak{T}$ in the equivalent IR $\alpha = + 1 / 2$, because of the Wigner criterion $W_{+ 1 / 2}^{\mathfrak{T}} = - 1$ \eqref{eq:194_Wigner_criterion_Mz_Au_T}.
On the other hand, since $W_{+ 1 / 2}^{\mathfrak{C}} = W_{+ 1 / 2}^\Gamma = 0$ \eqref{eq:194_Wigner_criterion_Mz_Au_C}, \eqref{eq:194_orthogonality_test_Mz_Au_G}, the lower left particle is mapped to the upper right holes in the nonequivalent IR $\alpha = - 1 / 2$, by the pseudo-PHS $\mathfrak{C}$ and the CS $\Gamma$.
Therefore, the $\alpha = + 1 / 2$ Hamiltonian block [the red frame in Fig.~\ref{fig:effective_AZ_class_Mz_Au}(b)] has the pseudo-TRS $\mathfrak{T}$ with $\mathfrak{T}^2 = - E$, and therefore belongs to the AZ class AII~\cite{Zirnbauer1996, Altland1997}.

As indicated in the above discussions, the EAZ class represents the property of a Hamiltonian block with respect to the onsite symmetries, where the block is obtained by block-diagonalizing the BdG Hamiltonian into eigenspaces (IRs) of the crystal symmetry.
In general, therefore, the EAZ class of a block is different from the AZ class of the whole Hamiltonian.

We have looked at an example of \textit{a nontrivial line node on a mirror plane stemming from nonsymmorphic symmetry} in terms of the topological classification.
The classification theory is of course applicable to other various crystal and/or magnetic symmetries.
Tables~\ref{tab:planes_classification_group_theory} and \ref{tab:planes_classification_topology} respectively show group-theoretical~\cite{Micklitz2017_PRB, Micklitz2017_PRL, Sumita2018} and topological~\cite{Kobayashi2018} classifications of gap structures in mirror- or glide-symmetric superconductors.
The classification results are categorized into six types (A)--(F), where (A) and (B) represent a FM superconductor without TRS, and (C)--(F) indicate a PM or AFM one with TRS.
Key ingredients of the classification are the $z$ components of $\bm{t}_\mathcal{M}$ and $\bm{t}_\mathcal{T}$, which are translation parts of the mirror (glide) operator $\mathcal{M}$ and the time-reversal operator $\mathcal{T}$, respectively.
A nonzero $[\bm{t}_\mathcal{M}]_z$ indicates the presence of screw symmetry along the $z$ axis in the superconductor.
When $[\bm{t}_\mathcal{T}]_z$ is nonzero, on the other hand, the superconductor is in an AFM state with a $z$-directional propagating vector.%
\footnote{We note that, even when $[\bm{t}_\mathcal{T}]_z$ is zero, an AFM state with an $x$- or $y$-directional propagating vector as well as a PM state is allowed.}
For example, the magnetic space group $M = P6_3/mmc1'$ discussed above is categorized into the case (E) since it is PM and screw-symmetric.
In general, when $[\bm{t}_\mathcal{M}]_z$ or $[\bm{t}_\mathcal{T}]_z$ is nonzero, namely in the cases (B), (D), (E), and (F), superconducting gap structures on the basal plane $k_z = 0$ and the zone face $k_z = \pi$ are different (see Table~\ref{tab:planes_classification_group_theory}).
This is a consequence of nonsymmorphic symmetry.

\begin{table}[tbp]
 \caption{Group-theoretical classification of superconducting gaps on high-symmetry planes. The fourth and fifth columns indicate representations allowed for a gap function on the basal plane $k_z = 0$ and the zone face $k_z = \pi$, respectively. The representations are decomposed into the IRs of the point group $C_{2h}$. A gap function should be zero, and thus, a node appears, if the corresponding IR does not exist in the reductions~\cite{Izyumov1989, Yarzhemsky1992, Yarzhemsky1998, Yarzhemsky2000, Yarzhemsky2003, Yarzhemsky2008}; otherwise, the superconducting gap opens in general. Adapted with permission from Refs.~\cite{Sumita2018, Kobayashi2018}. Copyright \copyright{} 2018 by the American Physical Society.}
 \label{tab:planes_classification_group_theory}
 \begin{center}
  \begin{tabular}{cclcc} \hline\hline
   Case & TRS exists? & \multicolumn{1}{c}{Key ingredients} & $k_z = 0$ & $k_z = \pi$ \\ \hline
   (A) & No  & $[\bm{t}_\mathcal{M}]_z = 0$ & $A_u$ & $A_u$ \\
   (B) & No  & $[\bm{t}_\mathcal{M}]_z \neq 0$ & $A_u$ & $B_u$ \\
   (C) & Yes & $[\bm{t}_\mathcal{M}]_z = [\bm{t}_\mathcal{T}]_z = 0$ & $A_g + 2 A_u + B_u$ & $A_g + 2 A_u + B_u$ \\
   (D) & Yes & $[\bm{t}_\mathcal{M}]_z = 0, [\bm{t}_\mathcal{T}]_z \neq 0$ & $A_g + 2 A_u + B_u$ & $B_g + 3 A_u$ \\
   (E) & Yes & $[\bm{t}_\mathcal{M}]_z \neq 0, [\bm{t}_\mathcal{T}]_z = 0$ & $A_g + 2 A_u + B_u$ & $A_g + 3 B_u$ \\
   (F) & Yes & $[\bm{t}_\mathcal{M}]_z \neq 0, [\bm{t}_\mathcal{T}]_z \neq 0$ & $A_g + 2 A_u + B_u$ & $B_g + A_u + 2 B_u$ \\ \hline\hline
  \end{tabular}
 \end{center}
\end{table}

\begin{table}[tbp]
 \centering
 \caption{Topological classification of superconducting gaps for the time-reversal symmetric superconductors labeled (C)--(F) in Table~\ref{tab:planes_classification_group_theory}. The table includes classification for only 0D topological invariants. Adapted with permission from Ref.~\cite{Kobayashi2018}. Copyright \copyright{} 2018 by the American Physical Society.}
 \label{tab:planes_classification_topology}
 \begin{center}
  \begin{tabular}{ccccccccc} \hline\hline
   & \multicolumn{2}{c}{$A_g$} & \multicolumn{2}{c}{$B_g$} & \multicolumn{2}{c}{$A_u$} & \multicolumn{2}{c}{$B_u$} \\ \hline
   Case & $k_z = 0$ & $k_z = \pi$ & $k_z = 0$ & $k_z = \pi$ & $k_z = 0$ & $k_z = \pi$ & $k_z = 0$ & $k_z = \pi$ \\ \hline
   (C) & AIII, $0$ & AIII, $0$          & D, $\mathbb{Z}_2$ & D, $\mathbb{Z}_2$  & C, $0$ & C, $0$             & AIII, $0$ & AIII, $0$          \\
   (D) & AIII, $0$ & AII, $2\mathbb{Z}$ & D, $\mathbb{Z}_2$ & DIII, $0$          & C, $0$ & C, $0$             & AIII, $0$ & AII, $2\mathbb{Z}$ \\
   (E) & AIII, $0$ & DIII, $0$          & D, $\mathbb{Z}_2$ & AII, $2\mathbb{Z}$ & C, $0$ & AII, $2\mathbb{Z}$ & AIII, $0$ & C, $0$             \\
   (F) & AIII, $0$ & D, $\mathbb{Z}_2$  & D, $\mathbb{Z}_2$ & AIII, $0$          & C, $0$ & AIII, $0$          & AIII, $0$ & C, $0$             \\ \hline\hline
  \end{tabular}
 \end{center}
\end{table}

Comparing Tables~\ref{tab:planes_classification_group_theory} and \ref{tab:planes_classification_topology}, we find a one-to-one correspondence between the group-theoretical and topological classification theories; when the presence of a line node is predicted in the group-theoretical classification, there exists a 0D topological invariant characterizing the node.
Furthermore, we have revealed that a line node in an even-parity superconductor is protected by a 1D winding number as well as the 0D invariant~\cite{Kobayashi2018}, although this is not shown in the tables.
The presence of the winding number means that the node is more stable against fluctuations, and indicates the emergence of a Majorana flat band as a surface state.
For further details, see Ref.~\cite{Kobayashi2018}.

\subsection{Classification on high-symmetry lines: nontrivial gap structures due to angular momentum}
\label{sec:gap_classification_lines}
Next, let us move on to classification results on high-symmetry lines in the BZ.
In this subsection, only symmorphic and PM space groups are discussed for simplicity.
Nevertheless, unconventional superconducting gap structures sometimes appear on the rotation-symmetric axes because of \textit{angular momentum of a normal Bloch state}.%
\footnote{This does not mean that nonsymmorphic symmetry on the rotational axes is unimportant. Indeed, Ref.~\cite{Yoshida2019} found that gap structures on $C_{2v}$-symmetric lines sometimes become nontrivial due to glide symmetry.}

Table~\ref{tab:lines_classification_topology} shows topological gap classification (see Sect.~\ref{sec:gap_classification_topology}) on $n$-fold rotational symmetric lines in the BZ ($n = 2$, $3$, $4$, $6$)~\cite{Sumita2019}.
We emphasize that representations with \textit{higher-angular-momentum states}, such as $3 / 2$ and $5 / 2$ spins, appear for $n \geq 3$, while a normal Bloch state on a twofold rotational axis always complies with spin $1 / 2$.
Surprisingly, superconducting gap classification results on threefold and sixfold lines depend on the angular momentum of the normal Bloch state, although they are unique for twofold- and fourfold-symmetric cases.
The \textit{angular-momentum-dependent gap structures} are important findings in our modern classification theory~\cite{Sumita2018, Sumita2019, Sumita2020_kotaibutsuri, Sumita2021_book}, since they cannot be predicted by the conventional classification theory of order parameters~\cite{Volovik1984, Volovik1985, Anderson1984, Blount1985, Ueda1985, Sigrist-Ueda}.

\begin{table}[tbp]
 \caption{Topological gap classification on high-symmetry lines in the BZ. Each classification is characterized by the type of topological number and gap structure: (G) full gap, (P) point nodes, (L) line nodes, and (S) surface nodes (Bogoliubov Fermi surfaces). For details of point groups and their IRs, see Refs.~\cite{Inui-Tanabe-Onodera, Bilbao}. In a spontaneous TRS breaking phase, 2D IRs in $D_{3d}$, $D_{4h}$, and $D_{6h}$ are the same as those in $S_6$, $C_{4h}$, and $C_{6h}$, respectively, since all the 2D IRs are decomposed into 1D IRs with different eigenvalues of the rotation symmetry. Therefore we do not show the 2D IRs in tables (f), (g), and (h). Adapted with permission from Ref.~\cite{Sumita2019}. Copyright \copyright{} 2019 by the American Physical Society.}
 \label{tab:lines_classification_topology}
 \begin{center}
  \scalebox{0.95}{%
  \begin{tabular}{cccp{10mm}cccp{5mm}ccc} \hline\hline
   \multicolumn{3}{c}{(a) $\bar{\mathcal{G}}^{\bm{k}} = C_2$, $\alpha = \pm 1 / 2$} & & \multicolumn{3}{c}{(b1) $\bar{\mathcal{G}}^{\bm{k}} = C_3$, $\alpha = + [-] 1 / 2$} & \quad & \multicolumn{3}{c}{(b2) $\bar{\mathcal{G}}^{\bm{k}} = C_3$, $\alpha = \pm 3 / 2$} \\ \cline{1-3}\cline{5-7}\cline{9-11}
   IR of $C_{2h}$ & EAZ & Classification & & IR of $S_6$ & EAZ & Classification & & IR of $S_6$ & EAZ & Classification \\ \cline{1-3}\cline{5-11}
   $A_g$ & AIII & $0$ (G)            & & $A_g$               & AIII & $0$ (G)            & & $A_g$        & DIII & $0$ (G)          \\
   $A_u$ & AIII & $0$ (G)            & & $A_u$               & AIII & $0$ (G)            & & $A_u$        & CII  & $0$ (G)          \\
   $B_g$ & D    & $\mathbb{Z}_2$ (L) & & $^{2}E_g [^{1}E_g]$ & D    & $\mathbb{Z}_2$ (S) & & $^{1, 2}E_g$ & A    & $\mathbb{Z}$ (S) \\
   $B_u$ & C    & $0$ (G)            & & $^{1}E_g [^{2}E_g]$ & A    & $\mathbb{Z}$ (S)   & &              &      &                  \\
         &      &                    & & $^{2}E_u [^{1}E_u]$ & C    & $0$ (G)            & & $^{1, 2}E_u$ & A    & $\mathbb{Z}$ (P) \\
         &      &                    & & $^{1}E_u [^{2}E_u]$ & A    & $\mathbb{Z}$ (P)   & &              &      &                  \\
   \\
   \multicolumn{3}{c}{(c) $\bar{\mathcal{G}}^{\bm{k}} = C_4$, $\alpha = + [-] 1 / 2, + [-] 3 / 2$} & & \multicolumn{3}{c}{(d1) $\bar{\mathcal{G}}^{\bm{k}} = C_6$, $\alpha = + [-] 1 / 2, + [-] 5 / 2$} & & \multicolumn{3}{c}{(d2) $\bar{\mathcal{G}}^{\bm{k}} = C_6$, $\alpha = \pm 3 / 2$} \\ \cline{1-3}\cline{5-7}\cline{9-11}
  IR of $C_{4h}$ & EAZ & Classification & & IR of $C_{6h}$ & EAZ & Classification & & IR of $C_{6h}$ & EAZ & Classification \\ \cline{1-3}\cline{5-11}
   $A_g$               & AIII & $0$ (G)            & & $A_g$                     & AIII & $0$ (G)            & & $A_g$           & AIII & $0$ (G)            \\
   $A_u$               & AIII & $0$ (G)            & & $A_u$                     & AIII & $0$ (G)            & & $A_u$           & AIII & $0$ (G)            \\
   $B_g$               & A    & $\mathbb{Z}$ (L)   & & $B_g$                     & A    & $\mathbb{Z}$ (L)   & & $B_g$           & D    & $\mathbb{Z}_2$ (L) \\
   $B_u$               & A    & $\mathbb{Z}$ (P)   & & $B_u$                     & A    & $\mathbb{Z}$ (P)   & & $B_u$           & C    & $0$ (G)            \\
   $^{2}E_g [^{1}E_g]$ & D    & $\mathbb{Z}_2$ (S) & & $^{1}E_{1g} [^{2}E_{1g}]$ & D    & $\mathbb{Z}_2$ (S) & & $^{1, 2}E_{1g}$ & A    & $\mathbb{Z}$ (S)   \\
   $^{1}E_g [^{2}E_g]$ & A    & $\mathbb{Z}$ (S)   & & $^{2}E_{1g} [^{1}E_{1g}]$ & A    & $\mathbb{Z}$ (S)   & &                 &      &                    \\
   $^{2}E_u [^{1}E_u]$ & C    & $0$ (G)            & & $^{1}E_{1u} [^{2}E_{1u}]$ & C    & $0$ (G)            & & $^{1, 2}E_{1u}$ & A    & $\mathbb{Z}$ (P)   \\
   $^{1}E_u [^{2}E_u]$ & A    & $\mathbb{Z}$ (P)   & & $^{2}E_{1u} [^{1}E_{1u}]$ & A    & $\mathbb{Z}$ (P)   & &                 &      &                    \\
                       &      &                    & & $^{1, 2}E_{2g}$           & A    & $\mathbb{Z}$ (S)   & & $^{1, 2}E_{2g}$ & A    & $\mathbb{Z}$ (S)   \\
                       &      &                    & & $^{1, 2}E_{2u}$           & A    & $\mathbb{Z}$ (P)   & & $^{1, 2}E_{2u}$ & A    & $\mathbb{Z}$ (P)   \\
   \\
   \multicolumn{3}{c}{(e) $\bar{\mathcal{G}}^{\bm{k}} = C_{2v}$, $\alpha = 1 / 2$} & & \multicolumn{3}{c}{(f1) $\bar{\mathcal{G}}^{\bm{k}} = C_{3v}$, $\alpha =  1 / 2$} & \quad & \multicolumn{3}{c}{(f2) $\bar{\mathcal{G}}^{\bm{k}} = C_{3v}$, $\alpha = 3 / 2$} \\ \cline{1-3}\cline{5-7}\cline{9-11}
   IR of $D_{2h}$ & EAZ & Classification & & IR of $D_{3d}$ & EAZ & Classification & & IR of $D_{3d}$ & EAZ & Classification \\ \cline{1-3}\cline{5-11}
   $A_g$    & CI  & $0$ (G)            & & $A_{1g}$ & CI       & $0$ (G)            & & $A_{1g}$ & AIII     & $0$ (G)            \\
   $A_u$    & CI  & $0$ (G)            & & $A_{1u}$ & CI       & $0$ (G)            & & $A_{1u}$ & C        & $0$ (G)            \\
   $B_{1g}$ & BDI & $\mathbb{Z}_2$ (L) & & $A_{2g}$ & BDI      & $\mathbb{Z}_2$ (L) & & $A_{2g}$ & D        & $\mathbb{Z}_2$ (L) \\
   $B_{1u}$ & BDI & $\mathbb{Z}_2$ (P) & & $A_{2u}$ & BDI      & $\mathbb{Z}_2$ (P) & & $A_{2u}$ & AIII     & $0$ (G)            \\
   $B_{2g}$ & BDI & $\mathbb{Z}_2$ (L) & & 2D IRs   & see (b1) &                    & & 2D IRs   & see (b2) &                    \\
   $B_{2u}$ & CI  & $0$ (G)            & &                                                                                       \\
   $B_{3g}$ & BDI & $\mathbb{Z}_2$ (L) & &                                                                                       \\
   $B_{3u}$ & CI  & $0$ (G)            & &                                                                                       \\
   \\
   \multicolumn{3}{c}{(g) $\bar{\mathcal{G}}^{\bm{k}} = C_{4v}$, $\alpha = 1 / 2, 3 / 2$} & & \multicolumn{3}{c}{(h1) $\bar{\mathcal{G}}^{\bm{k}} = C_{6v}$, $\alpha = 1 / 2, 5 / 2$} & & \multicolumn{3}{c}{(h2) $\bar{\mathcal{G}}^{\bm{k}} = C_{6v}$, $\alpha = 3 / 2$} \\ \cline{1-3}\cline{5-7}\cline{9-11}
   IR of $D_{4h}$ & EAZ & Classification & & IR of $D_{6h}$ & EAZ & Classification & & IR of $D_{6h}$ & EAZ & Classification \\ \cline{1-3}\cline{5-11}
   $A_{1g}$ & CI      & $0$ (G)            & & $A_{1g}$ & CI       & $0$ (G)            & & $A_{1g}$ & CI       & $0$ (G)            \\
   $A_{1u}$ & CI      & $0$ (G)            & & $A_{1u}$ & CI       & $0$ (G)            & & $A_{1u}$ & CI       & $0$ (G)            \\
   $A_{2g}$ & BDI     & $\mathbb{Z}_2$ (L) & & $A_{2g}$ & BDI      & $\mathbb{Z}_2$ (L) & & $A_{2g}$ & BDI      & $\mathbb{Z}_2$ (L) \\
   $A_{2u}$ & BDI     & $\mathbb{Z}_2$ (P) & & $A_{2u}$ & BDI      & $\mathbb{Z}_2$ (P) & & $A_{2u}$ & BDI      & $\mathbb{Z}_2$ (P) \\
   $B_{1g}$ & AI      & $\mathbb{Z}$ (L)   & & $B_{1g}$ & AI       & $\mathbb{Z}$ (L)   & & $B_{1g}$ & BDI      & $\mathbb{Z}_2$ (L) \\
   $B_{1u}$ & AI      & $\mathbb{Z}$ (P)   & & $B_{1u}$ & AI       & $\mathbb{Z}$ (P)   & & $B_{1u}$ & CI       & $0$ (G)            \\
   $B_{2g}$ & AI      & $\mathbb{Z}$ (L)   & & $B_{2g}$ & AI       & $\mathbb{Z}$ (L)   & & $B_{2g}$ & BDI      & $\mathbb{Z}_2$ (L) \\
   $B_{2u}$ & AI      & $\mathbb{Z}$ (P)   & & $B_{2u}$ & AI       & $\mathbb{Z}$ (P)   & & $B_{2u}$ & CI       & $0$ (G)            \\
   2D IRs   & see (c) &                    & & 2D IRs   & see (d1) &                    & & 2D IRs   & see (d2) &                    \\ \hline\hline
  \end{tabular}
  }
 \end{center}
\end{table}

\begin{figure}[tbp]
 \centering
 \includegraphics[width=14cm, clip]{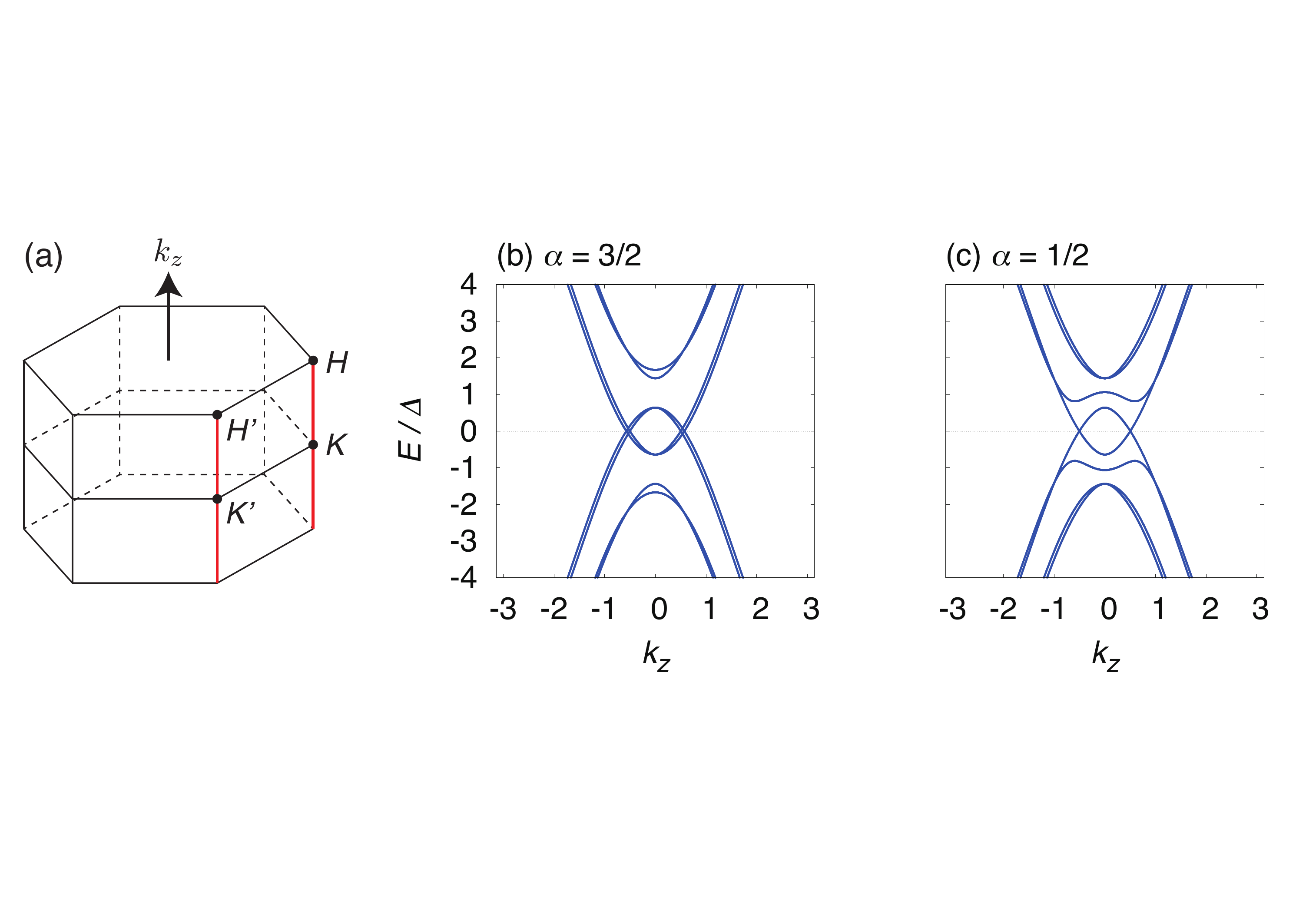}
 \caption{(a) A threefold-rotation-invariant axis $K$--$H$ in a hexagonal BZ. (b), (c) Quasiparticle energy spectra on the $K$--$H$ line in the TRS broken superconductor UPt$_3$~\cite{Sumita2019}. The normal Bloch state at the Fermi level has angular momenta $3 / 2$ and $1 / 2$ in (b) and (c), respectively. Adapted with permission from Refs.~\cite{Sumita2018, Sumita2019}. Copyright \copyright{} 2018-2019 by the American Physical Society.}
 \label{fig:UPt3_eigenKH_eta1}
\end{figure}

On $C_3$-symmetric lines, e.g., $^{1}E_u$ and $^{2}E_u$ superconducting order parameters become gapless and gapped for the $\alpha = + 1 / 2$ normal Bloch state, respectively [Table~\ref{tab:lines_classification_topology}(b1)], while both IRs are gapless for the $\alpha = + 3 / 2$ state [Table~\ref{tab:lines_classification_topology}(b2)].
Indeed, we have reproduced the result in an effective model of the hexagonal chiral superconductor UPt$_3$~\cite{Sumita2019}.
Figures~\ref{fig:UPt3_eigenKH_eta1}(b) and \ref{fig:UPt3_eigenKH_eta1}(c) show Bogoliubov quasiparticle spectra on the $C_3$-symmetric $K$-$H$ line in the BZ for $\alpha = 3 / 2$ and $\alpha = 1 / 2$, respectively.
Both bands generate point nodes on the line in the $\alpha = 3 / 2$ state [Fig.~\ref{fig:UPt3_eigenKH_eta1}(b)], whereas one band is gapless but the other is gapped for $\alpha = 1 / 2$ [Fig.~\ref{fig:UPt3_eigenKH_eta1}(c)].
These results are consistent with the above classification.
Furthermore, there are other candidate superconductors for unconventional angular-momentum-dependent gap structures; for details, see Refs.~\cite{Sumita2018, Sumita2019, Sumita2020_kotaibutsuri, Sumita2021_book}.
One can identify superconducting gap structures in various superconductors, by using our modern classification theory.

\section{Summary}
\label{sec:summary}
In this paper, we have reviewed the modern topological classification theory of superconducting gaps, and introduced various nontrivial nodal structures discovered by the theory.
By considering a specific $\bm{k}$ point in the BZ, our modern theory can take into account detailed properties such as nonsymmorphic symmetry and higher-spin states, which are not included in the conventional classification theory of order parameters~\cite{Volovik1984, Volovik1985, Anderson1984, Blount1985, Ueda1985, Sigrist-Ueda}.
Through comprehensive classification on high-symmetry planes and lines, the following important results are obtained:
\begin{itemize}
 \item difference of gap structures between a basal plane and a zone face attributed to \textit{nonsymmorphic symmetry}, and 
 \item gap structures depending on the \textit{angular momentum} of the normal Bloch state.
\end{itemize}
The classification results are determined only by symmetry, and therefore are universal and applicable to many candidate superconductors.

Finally, we mention more recent developments concerning the study.
The first example is the uranium-based superconductor UTe$_2$ discovered at the end of 2018~\cite{Ran2019}.
Since many previous experiments support intrinsic spin-triplet superconductivity in UTe$_2$, research into this material has developed explosively.
Stimulated by this background, we have discussed a detailed gap classification and the possibility of topological superconductivity, based on band structures obtained from first-principles calculations~\cite{Ishizuka2019}.
Furthermore, we have suggested unconventional gap structures protected by nonsymmorphic symmetry in LaNiGa$_2$~\cite{Badger2021_arXiv}, by using our classification method.
Like the above cases, our classification method is easily applicable to various superconductors in a systematic way.
In addition, more recent studies have proposed a more thorough classification of topological crystalline nodes and databases of nodal superconductors~\cite{Ono2021_arXiv, Tang2021_arXiv}.
Based on our theory and the related studies, many other superconductors hosting unconventional nodes should be discovered.

\section*{Acknowledgements}
The authors are grateful to Takuya Nomoto, Shingo Kobayashi, Ken Shiozaki, and Masatoshi Sato for fruitful discussions and collaborations on related topics.
This work was supported by JST CREST Grant No. JPMJCR19T2, JSPS KAKENHI (Grants No. JP18H05227, No. JP18H01178, No. JP20H05159), and SPIRITS 2020 of Kyoto University.

\appendix
\section{Wigner criterion}
\label{sec:Wigner_criterion}
In this appendix, we review the method and meaning of the Wigner criterion~\cite{Wigner, Herring1937, Inui-Tanabe-Onodera, Bradley-Cracknell, Shiozaki2018_arXiv1} used in the main text.
Let $\mathfrak{G}$ be a unitary group and $\alpha$ a $d_\alpha$-dimensional IR of $\mathfrak{G}$.
We choose a certain set of basis functions $\{\psi_1, \dots, \psi_{d_\alpha}\}$ of the IR $\alpha$.
$\psi_i$ transforms under the symmetry operation $g \in \mathfrak{G}$ as
\begin{equation}
 g \psi_i = \sum_{j = 1}^{d_\alpha} \psi_j [\Delta_\alpha(g)]_{ji},
\end{equation}
where $\Delta_\alpha$ is a representation matrix of the IR $\alpha$.

Let us consider whether the degeneracy of the IR ($d_\alpha$) increases or not, by adding an antiunitary operator $a$ to the group $\mathfrak{G}$.
In other words, we investigate the dimension of the representation matrix for the nonunitary group $\mathfrak{M} = \mathfrak{G} + a \mathfrak{G}$, supposing that information on the factor system $\{\omega(m_1, m_2)\} \in H^2(\mathfrak{M}, U(1)_\phi)$ is given.
The problem can be solved by the Wigner criterion:
\begin{equation}
 W_\alpha^a \equiv \frac{1}{|\mathfrak{G}|} \sum_{g \in \mathfrak{G}} \omega(a g, a g) \chi[\Delta_\alpha((a g)^2)] =
  \begin{cases}
   1, & \text{(a)}, \\
   -1, & \text{(b)}, \\
   0, & \text{(c)}.
  \end{cases}
\end{equation}
For each case, indeed, the irreducible corepresentation matrix $D_\alpha$ of $\mathfrak{M}$, which corresponds to $\Delta_\alpha$ of $\mathfrak{G}$, is constructed as follows~\cite{Wigner, Bradley-Cracknell, Inui-Tanabe-Onodera, Shiozaki2018_arXiv1},
\begin{align}
 D_\alpha(g) &=
 \begin{cases}
  \Delta_\alpha(g), & \text{(a)}, \\
  \begin{pmatrix}
   \Delta_\alpha(g) & 0 \\
   0 & \Delta_\alpha(g) 
  \end{pmatrix}, & \text{(b)}, \\
  \begin{pmatrix}
   \Delta_\alpha(g) & 0 \\
   0 & \frac{\omega(g, a)}{\omega(a, a^{-1} g a)} \Delta_\alpha(a^{-1} g a)^*
  \end{pmatrix}, & \text{(c)},
 \end{cases} \\
 D_\alpha(a) &=
 \begin{cases}
  U, & \text{(a)}, \\
  \begin{pmatrix}
   0 & -U \\
   U & 0
  \end{pmatrix}, & \text{(b)}, \\
  \begin{pmatrix}
   0 & \omega(a, a) \Delta_\alpha(a^2) \\
   1 & 0
  \end{pmatrix}, & \text{(c)},
 \end{cases}
\end{align}
where $U$ used in the (a) and (b) cases is a unitary matrix satisfying
\begin{equation}
 U^\dagger \Delta_\alpha(g) U = \frac{\omega(g, a)}{\omega(a, a^{-1} g a)} \Delta_\alpha(a^{-1} g a)^*.
\end{equation}
The meaning of each case is shown in the following.
\begin{enumerate}
 \renewcommand{\labelenumi}{(\alph{enumi})}
 \item There is no additional degeneracy due to the presence of the antiunitary operator $a$, because $\{\psi_i\}$ and $\{a \psi_i\}$ are not independent.
 \item The presence of the operator $a$ gives rise to additional degeneracy, because $\{a \psi_i\}$ is linearly independent of $\{\psi_i\}$ although they belong to the same IR $\alpha$.
 \item The degeneracy is doubled by applying $a$, because the basis $\{a \psi_i\}$ belongs to a representation $\alpha'$ inequivalent to $\alpha$.
\end{enumerate}

%

\end{document}